\documentclass[a4paper]{llncs}

\usepackage{url}
\usepackage{wrapfig}
\usepackage{subfigure}
\usepackage{epsfig}
\usepackage{calc}
\usepackage{amssymb}
\usepackage{amstext}
\usepackage{amsmath}
\usepackage{multicol}
\usepackage{pslatex}
\usepackage{pgfplots}
\usepackage{xspace}
\usepackage{multirow}
\usepackage{tablefootnote}
\usepackage{enumitem}
\setlist{topsep=0pt,itemsep=-1mm,partopsep=1mm,parsep=2mm}

\newcommand{\secbeg}{\vspace{-5mm}}
\newcommand{\secend}{\vspace{-3mm}}

\newcommand{\subsecbeg}{\vspace{-4mm}}
\newcommand{\subsecend}{\vspace{-1.5mm}}

\newcommand{\tabbeg}{\vspace{-5mm}}
\newcommand{\tabend}{\vspace{-8mm}}

\newcommand{\figbeg}{\vspace{-4mm}}
\newcommand{\figend}{\vspace{-4mm}}

\usepackage{parcolumns}
\usepackage{textcomp}
\usepackage{tikz}
\usetikzlibrary{calc}
\usetikzlibrary{decorations.pathreplacing,calc}
\newcommand{\tikzmark}[1]{\tikz[overlay,remember picture] \node (#1) {};}

\makeatletter
\def\@copyrightspace{\relax}
\makeatother

\makeatletter
\def\hlinewd#1{%
  \noalign{\ifnum0=`}\fi\hrule \@height #1 \futurelet
   \reserved@a\@xhline}
\makeatother

\makeatletter

\makeatletter

\newcommand*{\AddNoteRightBrace}[4]{%
    \begin{tikzpicture}[overlay, remember picture]
        \draw [decoration={brace,amplitude=0.3em},decorate, thick,black]
            ($(#1)!([yshift=1.5ex]#3)!($(#1)-(0,1)$)$) --  
            ($(#1)!(#2)!($(#1)-(0,1)$)$)
                node [below=2pt, align=center, text width=0.5cm, pos=0.5, anchor=west] {#4};
    \end{tikzpicture}
}%

\newcommand*{\AddNoteLeftBrace}[4]{%
    \begin{tikzpicture}[overlay, remember picture]
        \draw [decoration={brace,amplitude=0.3em},decorate, thick,black]
            ($(#1)!(#2)!($(#1)-(0,1)$)$) --  
            ($(#1)!([yshift=1.5ex]#3)!($(#1)-(0,1)$)$)
            
                node [above=2pt, align=center, text width=0.4cm, pos=0.5, anchor=east] {#4};
    \end{tikzpicture}
}%

\usepackage[normalem]{ulem}
\usepackage{soul}
\usepackage{listings}

\let\othelstnumber=\thelstnumber
\def\createlinenumber#1#2{
    \edef\thelstnumber{%
        \unexpanded{%
            \ifnum#1=\value{lstnumber}\relax
              #2%
            \else}%
        \expandafter\unexpanded\expandafter{\thelstnumber\othelstnumber\fi}%
    }
    \ifx\othelstnumber=\relax\else
      \let\othelstnumber\relax
    \fi
}

\usepackage[font=bf,skip=\baselineskip]{caption}

\lstdefinelanguage{ISA}{
  keywords={stw,ldw,entsp,lss, ldc, bf, bu, sub, bl, mul, retsp, mkmsk}
}[keywords]

\lstdefinelanguage{XC}{
  keywords={int, return, if}
}[keywords]

\lstdefinelanguage{hcir}{
  keywords={regtype, trust, num, list, rsize, var, resource,is,pred,check,true,checked,false, comp, intervals,int}
}[keywords]

\definecolor{light-gray}{gray}{0.80}

\newcommand{\level}{level\xspace}

\newcommand{\hcir}{HC IR\xspace}

\newcommand{\ciaopp}{CiaoPP\xspace}

\newcommand{\xmos}{XMOS\xspace}

\newcommand{\databranching}{\textbf{DDBr}}
\newcommand{\blockmodelHW}{\textbf{vs. HW}}

\urldef{\mailsa}\path|{umer.liqat,zorana.bankovic,pedro.lopez,manuel.hermenegildo}@imdea.org|
\urldef{\mailsb}\path|bueno@fi.upm.es|

\begin{document}

\title{Inferring Energy Bounds via Static Program Analysis and
  Evolutionary Modeling of Basic Blocks}

\author{
  Umer Liqat$^{1,3}$, Zorana Bankovi\'c$^1$, Pedro L\'{o}pez-Garc\'{i}a$^{1,2}$ \and  M.V.~Hermenegildo$^{1,3}$
}

\institute{$^1$ IMDEA Software Institute, Madrid, Spain \\ 
           $^2$ Spanish Council for Scientific Research (CSIC), Spain \\
           $^3$ Universidad Polit\'{e}cnica de Madrid, Spain\\
\mailsa
}

\maketitle

\begin{abstract}
The ever increasing number and complexity of energy-bound devices 
(such as the ones used in \emph{Internet of Things} applications,
smart phones, and mission critical systems) pose an important
challenge on techniques to optimize their energy consumption and
to verify that they will perform their function within the available
energy budget.  In this work we address this challenge from the
software point of view and propose a novel parametric approach to 
estimating tight bounds on the energy consumed by program executions
that are practical for their application to energy verification and
optimization.
Our approach divides a program into basic (branchless) blocks and
estimates the maximal and minimal energy consumption for each block
using an evolutionary algorithm.  Then it combines the obtained values
according to the program control flow, using static analysis, to infer
functions that give both upper and lower bounds on the energy
consumption of the whole program and its procedures as functions on
input data sizes.
We have tested our approach on (C-like) embedded programs running on
the \xmos hardware platform. However, our method is general enough to be
applied to other microprocessor architectures and programming
languages. The bounds obtained by our prototype implementation can be
tight while remaining on the safe side of budgets in practice, as
shown by our experimental evaluation.
\end{abstract}


\secbeg
\section{Introduction}
\label{sec:introduction}
\secend

Reducing and controlling the energy consumption and the environmental
impact of computing technologies has become a 
challenging problem worldwide. It is a significant issue in systems
ranging from small \emph{Internet of Things (IoT)} devices, sensors,
smart watches, smart phones and portable/implantable medical devices, to
large data centers and high-performance computing systems.

Trend analyses of the so called \emph{Internet of Things} paradigm
estimate that by the year 2020, about 50 billion small autonomous
devices, embedded in all kind of objects, even in our clothes or stuck
to our bodies, will operate and intercommunicate continuously for long
periods of time, such as years. Such devices rely on small batteries
or energy harvested from the environment, which implies that their
energy consumption should be very low.  Although there have been
improvements in battery and energy harvesting technology, they alone
are often not enough to achieve the required level of energy
consumption to fully support \emph{IoT} and other energy-bound
applications (e.g., sensor-based or signal-processing applications).
In addition, for many of these IoT and other applications (e.g., space
systems or implantable/portable medical devices), beyond optimizing
energy consumption, it is actually crucial to guarantee that execution
will complete within a specified energy budget, i.e., before the
available system energy runs out, or that the system will function for
at least a given period of time.

As mentioned before, 
energy consumption is also an issue at the large scale: as a result of
the huge growth in cloud computing, Internet traffic, high-performance
computing, and distributed applications, current data centers consume
very large amounts of energy, not only to process and transport data,
but also for cooling.

In spite of the recent rapid advances in energy-efficient hardware,
it is software that controls the hardware, so that far more
energy savings remain to be tapped by improving the software that runs
on these devices.

In this work we address the challenge from the software point of view,
focusing on the \emph{static} estimation of the energy consumed by
program executions (i.e., at compile time, without actually running
the programs with concrete data), as a 
basis for
energy optimization and verification.
Such estimations are given as functions on input data sizes, since
data sizes typically influence the energy consumed by a program, but
are not known at compile time.  This approach allows abstracting away
such sizes and inferring energy consumption in a way that is
parametric on them.

Different types of resource usage estimations are possible, such as,
e.g., probabilistic, average, or safe bounds.  However, not all types
of estimations are valid or useful for a given application. For
example, in order to verify/certify energy budgets,
\emph{safe upper and lower bounds} on energy
consumption are
required~\cite{energy-verification-hip3es2015-short,resource-verif-2012-short}.
Unfortunately, current approaches that guarantee that the bounds are
always safe tend to compromise their tightness seriously, inferring
overly conservative bounds, which are not useful in practice. With
this safety/tightness trade-off in mind, our goal is the development
of an analysis that infers tight bounds that are on the safe side in
most cases, in order to be practical for verification applications, as
well as for energy optimization.

Describing how energy \emph{verification} is performed is out of the
scope of this paper, and we refer the reader
to~\cite{resource-verif-iclp2010-short,resource-verif-2012-short} for
a detailed description on how upper and lower bounds on resource
usage in general can be used for verification within the \ciaopp
system~\cite{ciaopp-sas03-journal-scp-short}, and
to~\cite{energy-verification-hip3es2015-short} for a
specialization to energy consumption verification. 
Herein we focus instead on the \emph{inference} of energy bounds.
Nevertheless, in the following we provide the 
intuition on how these bounds 
are used in our system for verification and
certification:
assume that $E_{l}$ and $E_{u}$ are a lower and an upper bound
(respectively) on energy consumption inferred by our combined
modeling-analysis approach for a program, and that $E_{b}$ is an
energy budget expressed by a program specification, e.g., defined by
the capacity of the battery. Then:

\begin{enumerate}

\item If $E_{u} \leq E_{b}$, then the given program can be safely
  executed within the existing energy budget.

\item If $E_{l} \leq E_{b} \leq E_{u}$, it might be possible to
  complete the execution of the program, but we cannot claim it for
  certain.

\item If $E_{b} < E_{l}$, then it is not possible to execute the
  program (the system will run out of batteries before program
  execution is completed).
\end{enumerate}

Of the small number of static energy analyses proposed to date, only a
few~\cite{NMHLFM08-tooshort,isa-energy-lopstr13-final-short,isa-vs-llvm-fopara-short}
use
resource analysis frameworks that are aimed at inferring safe upper
and lower bounds on the resources used by program executions.
A crucial component in order for such frameworks to infer information
regarding hardware-dependent resources, and, in particular, energy, is a
low-level resource usage model, such as, e.g., a model of the energy
consumption of individual instructions. 
Examples of such instruction-level models are ~\cite{LL07}, at the
Java bytecode level, or ~\cite{Kerrison13-short}, at the Instruction
Set Architecture (ISA) level.

Clearly, the accuracy of the
bounds inferred by analysis depends on the nature and accuracy of
the low-level models.  Unfortunately, instruction-level models such as~\cite{LL07,Kerrison13-short} 
provide \emph{average} energy consumption values or functions, which
are not really suitable for safe upper- or lower-bounds analysis.
Furthermore, trying to obtain instruction-level models that provide
strict safe energy bounds would result in very conservative bounds.
Although when supplied with such models the static analysis would infer
high-level energy consumption functions providing strictly safe
bounds, these bounds would not be useful in general because of their
large inaccuracy.
For this reason, the analyses
in~\cite{NMHLFM08-tooshort,isa-energy-lopstr13-final-short,isa-vs-llvm-fopara-short}
used instead the already mentioned
instruction level average energy
models~\cite{LL07,Kerrison13-short}. However, this meant that the energy
functions inferred for the whole program
were not strict bounds, but rather approximations of the actual
bounds, and could possibly be below or above. 
This trade-off between safety and accuracy
is a major challenge in energy analysis.
In this paper we address this challenge by finding a good compromise
and providing a technique for the generation of
lower-level energy models which are
useful and effective in practice for
verification-type applications.

The main source of inaccuracy
in current instruction-level energy models is inter-instruction
dependence (including also data dependence),
which is not captured by most
models.  
On the other hand, the concrete sequences of instructions that appear
in programs exhibit worst cases that are not as pessimistic as
considering the worst case for each of the individual intervening
instructions. Based on this, we decided to use \emph{branchless
  blocks} of 
ISA 
instructions as the modeling unit instead of
individual instructions.  We divide the 
(ISA)
program into such \textit{basic blocks},
each a straight-line code sequence with exactly one entry to the block
(the first instruction) and one exit from the block (the last
instruction). We then measure the energy consumption of these basic
blocks, and determine a maximum (resp. minimum)
energy consumption for each block.  In this way the inter-instruction
data dependence discussed above and other factors are accounted for 
within each block.
The inter-instruction dependencies between blocks are still modeled in a
conservative way, and hence can be one of the sources of inaccuracy.
However, such modeling does not affect the correctness of the energy
bounds.
The energy values obtained for each block are supplied to our static
resource analysis, which combines them according to the program
control flow and produces functions that give both upper and
lower bounds on the energy consumption of the whole program and its
procedures as functions on input data sizes.

In order to find the maximum and minimum energy consumption of each
basic block we use an evolutionary algorithm (EA), varying the basic
block's input values and taking energy measurements directly from the
hardware for each input combination.  This way, we take advantage of
the fast
search space exploration provided by EAs.
The approach in~\cite{pallister2015data} 
also uses EAs for estimating worst case energy consumption. However,
it is applied to \emph{whole} programs, rather than at the basic block level.
A major disadvantage of such an approach is that, if there are
data-dependent branches in the programs,  
as is often the case,
the EA quickly loses accuracy, and does not converge
since different input combinations can trigger different sets of
instructions~\cite{pallister2015data}. This can make the problem
intractable.  In contrast, our approach combines EAs and static
analysis techniques in order to get the best of both worlds.
Our approach takes out the treatment of data-dependent branches from
the EA, so that the same sequence of instructions is always executed
in each basic block.  This way, the EA converges and estimates the
worst (resp. best) case energy of the basic blocks
with higher accuracy.
We take care of the program control flow dependencies by using static
analysis instead.

For concreteness, in our experiments we focus on the energy analysis
of programs written in XC~\cite{Watt2009}, running on the XS1-L
architecture~\cite{XS1-Architecture}, designed by
\xmos.\footnote{\url{http://www.xmos.com/}} However, our approach is
general enough to be applied as well to the analysis of other
architectures and other programming
languages and their associated lower-\level program representations.
XC is a high-level, C-based programming
language that includes extensions for concurrency, communication,
input/output operations, and real-time behavior.  Our experimental
setup infers energy consumption information by processing the ISA
(Instruction Set Architecture) code compiled from XC, and reflects it
up to the source code \level.  Such information is provided in the
form of \emph{functions on input data sizes}, and is expressed by
means of
\emph{assertions}~\cite{hermenegildo11:ciao-design-tplp-short}.

The results of our experiments suggest that our approach is quite
accurate,  
in the sense that the inferred energy bounds are close to the actual
maximum and minimum energy consumptions.
Furthermore, the energy estimations produced by our approach were
always safe, in the sense that they over-approximated the actual
bounds (i.e., the inferred upper bounds were above the actual
highest energy consumptions and the inferred lower bounds below the
actual
lowest energy consumptions).
We argue thus that our analysis provides a good practical compromise.

In summary, the main contributions of this paper are:

\begin{itemize}

\item A novel approach that combines dynamic and static analysis
  techniques for inferring tighter upper and lower bounds on the
  energy consumption of program executions as functions of input data
  sizes. The dynamic part is based on EAs, and produces low-level
  energy models that contain \emph{upper and lower bounds} on the cost
  of the elementary operations, as opposed to just average values.

\item The proposal of a new abstraction level at which to perform the
  energy modeling of program components, namely at the level of basic
  (branchless) blocks of ISA instructions, and a method based on EAs
  to dynamically (i.e., by profiling) obtain accurate and practical
  upper and lower bounds on the energy of such basic blocks, with a
  good safety/accuracy
  compromise.

\item A prototype implementation and experimental study that
  supports our claims.

\end{itemize}

In the rest of the paper, Section~\ref{sec:energymodel} explains our
technique for energy modeling of program basic blocks.
Section~\ref{sec:static} shows how these models are used by the static
analysis to infer upper and lower bounds on the energy consumed by
programs as functions of their input data
sizes. Section~\ref{sec:experiments} reports on an experimental
evaluation of our approach. Related work is discussed in
Section~\ref{sec:related-work}, and finally
Section~\ref{sec:conclusion} summarizes our conclusions.  
This work is an extended and improved version of the workshop
paper~\cite{basic-block-energy-hip3es2016}.

\secbeg
\section{Modeling the Energy Consumption of Blocks}
\label{sec:energymodel}
\secend

As mentioned before, the first step of our energy bounds analysis is
to determine upper and lower bounds on the energy consumption of each
basic (branchless) program block. We perform the modeling at this
level rather than at the instruction level in order to cater for
inter-instruction dependencies.  We first identify all the basic
blocks of the program, and then we perform a profiling of the energy
consumption of each of these blocks for different input data using an
EA. These steps are explained in the following sections.

\subsecbeg
\subsection{Identifying the Basic Blocks to be Modeled}
\subsecend

A \emph{basic block} over an inter-procedural control flow graph (CFG)
is a maximal sequence of distinct instructions, $S_1$ through $S_n$,
such that all instructions $S_k, 1<k<n$ have exactly one in-edge and
one out-edge~(excluding call/return edges), $S_1$ has one out-edge,
and $S_n$ has one in-edge.  A basic block therefore has exactly one
entry point at $S_1$ and one exit point at $S_n$.

In order to divide a program into such basic blocks, 
the program is first compiled to a lower-level representation, ISA in
our case.  A dataflow analysis of the ISA representation yields an
inter-procedural control flow graph (CFG).  A final control flow
analysis is carried out to infer basic blocks from the CFG.
These basic blocks are further modified so that they can be run and
their energy consumption measured independently by the EA.
Modifications for each basic block include:

\begin{enumerate}
\item A basic block with $k$ function call instructions is divided
  into $k+1$ basic blocks without the function call instructions.
\label{split}
\item A number of special ISA instructions (e.g., \emph{return},
  \emph{call}, \emph{entsp}) are omitted from the block. The cost of
  such instructions is measured separately and added to the cost of
  the block or the function.
\label{omit}
\item The harness function that runs the blocks in isolation provides
  the context to each block needed for the results to be applicable to
  the original program. For example the memory accesses in each block
  are transformed into accesses to a fixed address in the local memory
  of the harness function. The initial values placed in this local
  memory are the inputs to the block that the EA explores.
\label{memory}
\end{enumerate}

\captionsetup[lstlisting]{font={scriptsize, bf}}

\begin{figure}[t]
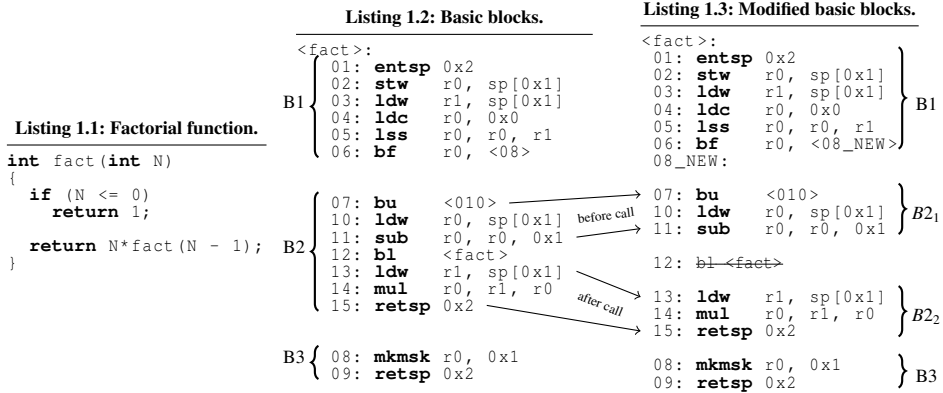

\figbeg \figbeg
\begin{parcolumns}[nofirstindent, colwidths={1=0.26\textwidth}]{3}
\colchunk{
\noindent\begin{minipage}{.28\textwidth}
\begin{lstlisting}[language={XC},caption={Factorial function.},frame=t, label={lst:factexample}]
int fact(int N) 
{
  if (N <= 0) 
    return 1;
  
  return N*fact(N - 1);
}
\end{lstlisting}
  \end{minipage}
} 
 \colchunk{
  \noindent\begin{minipage}{.32\textwidth}
\begin{lstlisting}[language={ISA},caption={Basic blocks.},frame=t, label={lst:basicblocks}]
<fact>:
  $\tikzmark{listing-b1-end}$ 01: entsp 0x2      
   02: stw   r0, sp[0x1]
   03: ldw   r1, sp[0x1]
   04: ldc   r0, 0x0
   05: lss   r0, r0, r1
  $\tikzmark{listing-b6-end}$ 06: bf    r0, <08> 
 
 
  $\tikzmark{listing-b7-end}$ 07: bu    <010> $\tikzmark{L1line7}$
   10: ldw   r0, sp[0x1]
   11: sub   r0, r0, 0x1 $\tikzmark{L1line11}$
   12: bl    <fact>
   13: ldw   r1, sp[0x1] $\tikzmark{L1line13}$
   14: mul   r0, r1, r0  
  $\tikzmark{listing-b15-end}$ 15: retsp 0x2 $\tikzmark{L1line15}$
 
 
  $\tikzmark{listing-b8-end}$ 08: mkmsk r0, 0x1
  $\tikzmark{listing-b9-end}$ 09: retsp 0x2

\end{lstlisting}
  \end{minipage}
 }
 \colchunk{
   \begin{minipage}{.30\textwidth}
\begin{lstlisting}[escapechar=!,language={ISA},caption={ Modified basic blocks.},frame=t, label={lst:basicblocksmodified}]
<fact>: 
 01: entsp 0x2 $\tikzmark{listing-4-end}$        $\tikzmark{listing-1-end}$
 02: stw   r0, sp[0x1]
 03: ldw   r1, sp[0x1]
 04: ldc   r0, 0x0
 05: lss   r0, r0, r1
 06: bf    r0, <08_NEW> $\tikzmark{listing-6-end}$
 08_NEW:
 
$\tikzmark{L2line7}$ 07: bu    <010>       $\tikzmark{listing-7-end}$
 10: ldw   r0, sp[0x1]
$\tikzmark{L2line11}$ 11: sub   r0, r0, 0x1 $\tikzmark{listing-11-end}$
 
 12: !\st{bl    <fact>}!
 
$\tikzmark{L2line13}$ 13: ldw   r1, sp[0x1] $\tikzmark{listing-13-end}$
 14: mul   r0, r1, r0
$\tikzmark{L2line15}$ 15: retsp 0x2 $\tikzmark{listing-15-end}$
 
 08: mkmsk r0, 0x1     $\tikzmark{listing-8-end}$
 09: retsp 0x2 $\tikzmark{listing-9-end}$
 
\end{lstlisting}
   \end{minipage}
 }
 \colplacechunks
\end{parcolumns}

\tikz[overlay,remember picture] \draw[color=black,->] ($(L1line7)+(0pt,0.5ex)$) -- ($(L2line7)+(0pt,0.5ex)$);

\tikz[overlay,remember picture] \draw[color=black,->] ($(L1line11)+(0pt,0.5ex)$) -- ($(L2line11)+(0pt,0.5ex)$) node [midway, sloped, above=1.8pt, fill=white] {\tiny before call};

\tikz[overlay,remember picture] \draw[color=black,->] ($(L1line13)+(0pt,0.5ex)$) -- ($(L2line13)+(0pt,0.5ex)$) node [midway, sloped, below=3pt, fill=white] {\tiny after call};

\tikz[overlay,remember picture] \draw[color=black,->] ($(L1line15)+(0pt,0.5ex)$) -- ($(L2line15)+(0pt,0.5ex)$);

\AddNoteRightBrace{listing-1-end}{listing-6-end}{listing-1-end}{\scriptsize B1}
\AddNoteRightBrace{listing-7-end}{listing-11-end}{listing-7-end}{\scriptsize $B2_1$}
\AddNoteRightBrace{listing-13-end}{listing-15-end}{listing-13-end}{\scriptsize $B2_2$}
\AddNoteRightBrace{listing-8-end}{listing-9-end}{listing-8-end}{\scriptsize B3}

\AddNoteLeftBrace{listing-b1-end}{listing-b6-end}{listing-b1-end}{\scriptsize B1}
\AddNoteLeftBrace{listing-b7-end}{listing-b15-end}{listing-b7-end}{\scriptsize B2}
\AddNoteLeftBrace{listing-b8-end}{listing-b9-end}{listing-b8-end}{\scriptsize B3}
\vspace{-24mm}
\caption{Example: Basic block modifications.}
\label{fig:blocks1}
\figend
\end{figure}

An example of modifications~\ref{split} and~\ref{omit} above is 
shown in Figure~\ref{fig:blocks1}, Listing~\ref{lst:basicblocks},
which is an ISA representation of a recursive factorial program where
the instructions are grouped together into 3 basic blocks $B1$,
$B2$, and $B3$.  Consider basic block $B2$. Since it has a
(recursive) function call to \emph{fact} at address 12, it is divided
further into two blocks in Listing~\ref{lst:basicblocksmodified}, such
that the instructions before and after the function call form two
blocks $B2_1$ and $B2_2$ respectively, and the call instruction ($bl$)
is omitted. The energy consumption of these two blocks is maximized
(minimized) by providing values to the input arguments to the block
(see below) using the EA.  The energy consumption of $B2$ can then be
characterized as:
\vspace{-3mm}
\begin{center}
$B2_e^A=B2_{1e}^A+B2_{2e}^A+bl_e^A$
\end{center}
\vspace{-1.5mm} where $B2_{1e}^A$, $B2_{2e}^A$, and $bl_e^A$ denote the
energy consumption of the $B2_1$, and $B2_2$ blocks, and the $bl$ ISA
instruction respectively, with approximation $A$ (where $A$=upper or
$A$=lower).

For each modified basic block, a set of input arguments is
inferred. This set is used for an individual representation to drive
the EA algorithm to maximize the energy consumption of the block. For
the entry block, the input arguments are derived from the signature of
the function. The set $gen(B)$ characterizes the set of variables read
without being previously defined in block $B$. It is defined as:

\vspace{-7mm}
\begin{align*}
gen(b)=\textstyle\bigcup\limits_{k=1}^{n}\{v\mid v\in \mathit{ref}(k) \wedge \forall (j<k). v\notin \mathit{def}(j)\}
\end{align*}
\vspace{-4mm}

\noindent where $\mathit{ref}(n)$ and $\mathit{def}(n)$ denote the variables
referred to and defined/updated at a node $n$ in block $b$, 
respectively.
For the basic blocks in 
Listing~\ref{lst:basicblocks} (Fig.~\ref{fig:blocks1}),
the input arguments are
$gen(B1)$=\{r0\}, $gen(B2_1)$=\{sp[0x1]\}, $gen(B2_2)$=\{sp[0x1],r0\},
and $gen(B3) = \emptyset$.

\subsecbeg

\subsection{Evolutionary Algorithm for finding Energy Bounds for Basic Blocks}

\subsecend

We now detail the
main aspects of the EA used
for estimating the maximum (i.e., worst case) and minimum (i.e., best
case)
energy consumption of a basic block. The only difference between the
two algorithms is the way we interpret the objective function: in the
first case we want to maximize it, while in the second one we want to
minimize it.\\ [-8mm]
\begin{wrapfigure}[12]{r}{0.25\textwidth}
\figbeg
\centering
\includegraphics[width=0.25\textwidth,scale=0.06]{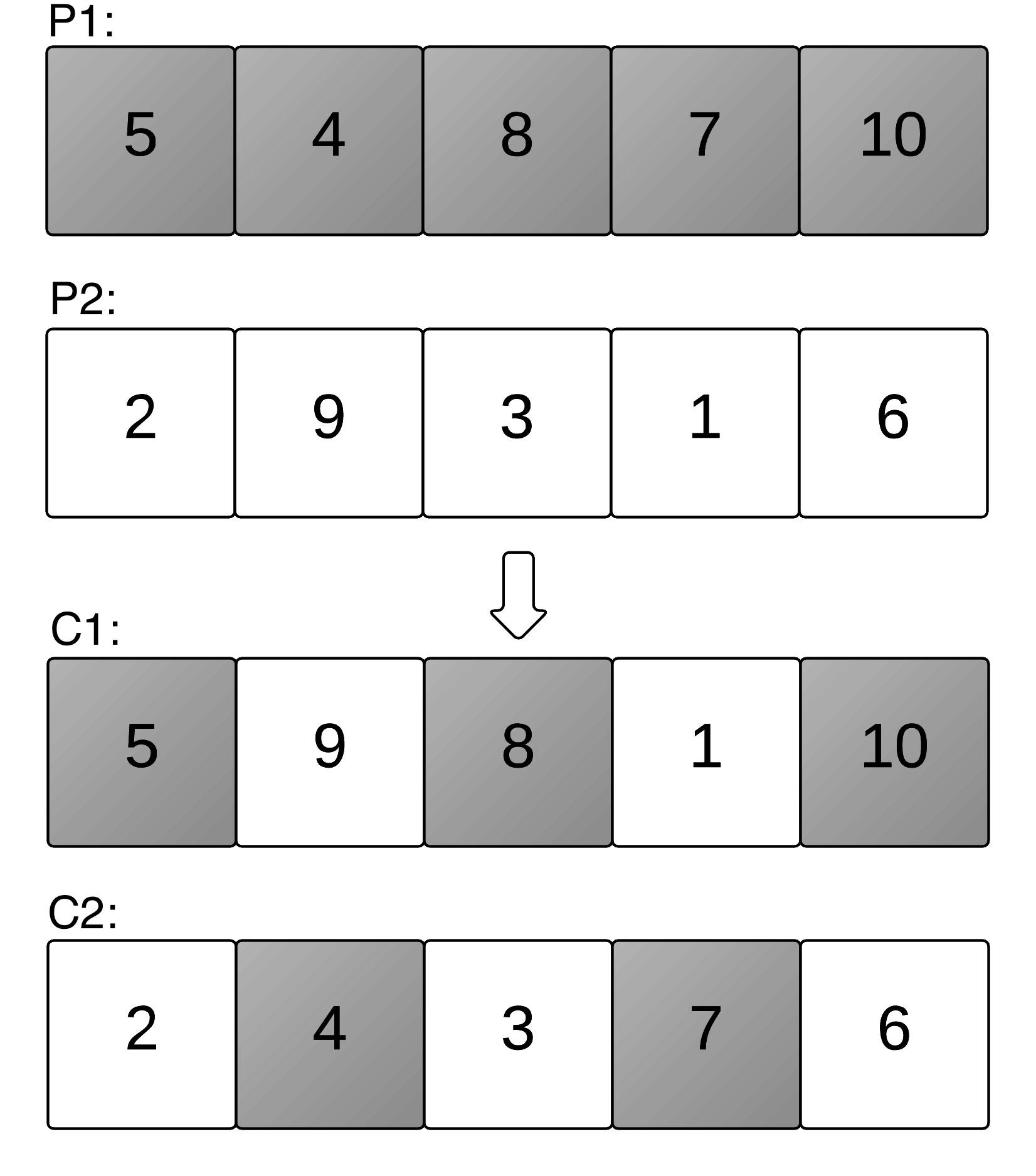}
\figend \figend
\caption{Crossover.}
\label{fig:indiv1}
\end{wrapfigure}

\vspace{.1cm}
\noindent
\textbf{Individual.} The search space dimensions are the different
input variables to the blocks. Our goal is to find the combination of
input values which maximizes (minimizes) the energy consumption of
each block. The set of input variables to a block is inferred using a
dataflow analysis (as explained in the previous section). Thus, an individual
is simply an array of input values given in the order of their
appearance in the block.
In the initial population, the input values to an individual are
randomly assigned to 32-bit numbers. In addition, some corner cases
that are known to cause high or low energy consumption for particular
instructions are included.\footnote{For example, all 1s for high energy
  consumption, or all 0s for low energy consumption as operands to a
  multiply ISA instruction.} 

\begin{wrapfigure}[7]{r}{0.4\textwidth}
\figbeg
\figbeg
\centering
\includegraphics[width=0.4\textwidth, scale=0.06]{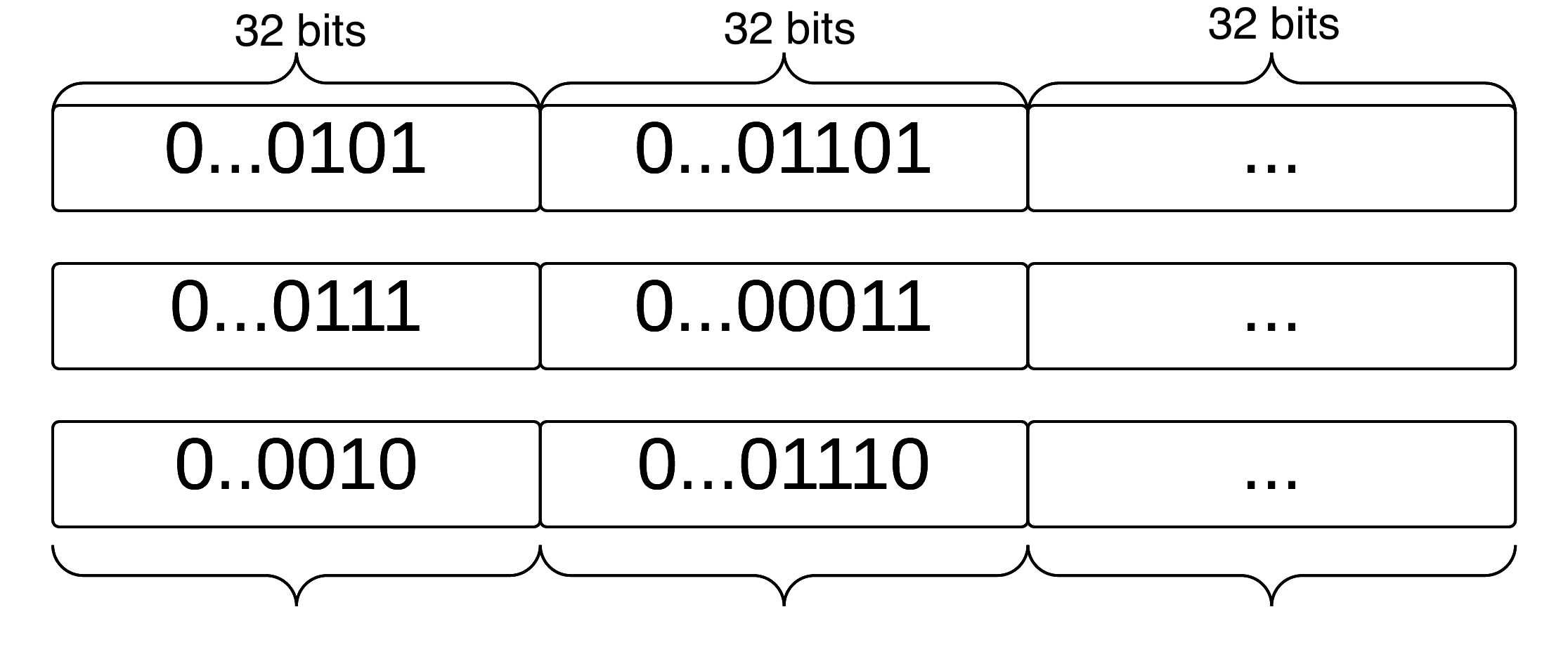}
\figend \figend
\caption{Mutation.}
\label{fig:mut1}
\figend
\end{wrapfigure}
\noindent
\textbf{Crossover.} The crossover operation is implemented as an even-odd
crossover, since it provides more variability than a standard $n$-point
crossover.
The process is depicted in Figure~\ref{fig:indiv1}, where $P1$ and
$P2$ are the parents, and $C1$ and $C2$ are their children created by
the crossover operation.

\vspace{.1cm}

\noindent
\textbf{Mutation.} For the purpose of this work we have created a
custom mutation operator. Since the energy consumption in digital
circuits is mainly the result of bit flipping, we believe that the
best way to explore the search space is by performing some bit
flipping in the mutation operation. This is implemented as
follows. For each gene component (i.e., for each input value to the basic block):

\begin{enumerate}
\item We create a random 32-bit integer (a random mask).
\item Then we perform the XOR operation of that integer and the
  corresponding gene. This results in a random flipping of the bits of
  each gene: only the bits of the gene at positions where the value of
  the random mask is $1$ are flipped.
\end{enumerate} 
The process is depicted in Figure~\ref{fig:mut1}, where the input
values are given as binary numbers.

In the ISA representation of the program, the type structure is
implicit and each operand (e.g., register) of an ISA instruction is a
32-bit value that either represents data or a memory address holding
data. Since the input variables to a block holds data (memory accesses
are transformed as described in the previous section), the mutation and
crossover operators
could generate data that 
such input variables would never take if the block were to run as part
of the whole program. Thus, this conservative modeling of inter-block
data dependencies could be one source of inaccuracy.

\vspace{.1cm}
\noindent
\textbf{Objective function.} The objective function that we want to
maximize/minimize is the energy of a basic block, which is measured
directly from the chip. The concrete measurement setting
will be explained in Section~\ref{sec:measurement-harness}.

In general, pipeline effects such as stalls (to resolve pipeline
hazards), which depend on the state of the processor at the start of
the execution of a basic block, can affect
the upper/lower bound estimated on the energy consumption of such a
block. Note that in our approach intra-block pipeline effects are
accounted 
for, since the dependencies among the instructions within a block are
captured.
However, the inter-block pipeline effects also need to be accounted
for. These can be modeled in a conservative way by assuming a maximum
stall penalty for the upper bound estimation of each block (e.g., by
adding a stall penalty to the execution time of the
block). Similarly, for the lower bound estimation a zero stall penalty
can be used. To approximate this effect,
in~\cite{Chakravarty-CODES-ISSS-13-short}, the authors characterize
each block through pairwise executions
with all of its possible predecessors.  Each basic block pair is
characterized by executing it on an Instruction Set Simulation (ISS)
to collect cycle counts. A similar reasoning would apply to cache
effects due to module boundaries.  These effects could also be bounded
using 
cache and pipeline analysis
techniques~\cite{wilhelm-cache-analysis-survey-2016-short}.
In any case, the \xmos XS1 architecture used in our experiments is a
\emph{cache-less, by design-predictable architecture}, and in
particular it does not exhibit these pipeline effects, 
since exactly one instruction per thread is executed in a 4-stage
pipeline (more details in can be found in Section~\ref{sec:xmos}).

\secbeg 
\section{Static Analysis of the Program Energy Consumption}
\label{sec:static}
\secend

Once energy models are obtained for each basic block of the program,
the energy consumption of the whole program is bounded by a static
analyzer that takes into account the control flow of the program and
infers safe upper/lower bounds on its energy consumption.
We have implemented such an analyzer by specializing the generic
resource analysis framework provided by
\ciaopp~\cite{plai-resources-iclp14-short} for programs written in
the XC programming language~\cite{Watt2009} and running on the \xmos
XS1-L architecture.  This includes the use of a
transformation~\cite{isa-energy-lopstr13-final-short,isa-vs-llvm-fopara-short}
of the ISA code into an intermediate representation for analysis
which is a series of connected code blocks, represented as Horn
Clauses (\hcir).
Such a transformation is shown in Fig.~\ref{fig:blocks-transform}
where the ISA representation of the factorial function from
Listing~\ref{lst:basicblocks} (Fig.~\ref{fig:blocks1}) is shown.  It
transforms the blocks into clauses and instructions into clause
literals. Conditional branching is modeled by predicates with two
clauses, one with the condition true and the other false. The
input/output arguments of each block are inferred via a dataflow
analysis. The final step transforms the blocks into Static Single
Assignment (SSA) form where each variable is assigned exactly once.
The analyzer deals with this \hcir always in the same way,
independently of
its origin, setting up cost equations for
all code blocks (predicates).  We have also written the necessary code
(i.e., assertions~\cite{hermenegildo11:ciao-design-tplp-short}) to
feed such analyzer with the block-level upper/lower bound energy model
obtained by using the technique explained in
Section~\ref{sec:energymodel}.
The analyzer
enables a programmer to symbolically bound the energy consumption of a
program $P$ on input data $\bar{x}$ without actually running
$P(\bar{x})$. It automatically sets up a system of recurrence (cost)
equations
that capture the cost (energy consumption) of $P$ as a function of the
sizes of its input arguments $\bar{x}$. Typical metrics used for data
sizes in this context are the actual value of a number, the length of
a list or array,
etc.~\cite{resource-iclp07-short,plai-resources-iclp14-short}.

\begin{figure}[ht]
\figbeg
\noindent\begin{minipage}{.38\textwidth}
\begin{lstlisting}[language=ISA, frame=tlrb, numbers=left,  numberblanklines=false]
<fact>:
01: entsp 0x2	
02: stw   r0, sp[0x1]
03: ldw   r1, sp[0x1]
04: ldc   r0, 0x0
05: lss   r0, r0, r1
06: bf    r0, <008>



07: bu    <010>  
10: ldw   r0, sp[0x1]
11: sub   r0, r0, 0x1
12: bl    <fact>

13: ldw   r1, sp[0x1]
14: mul   r0, r1, r0
15: retsp 0x2


08: mkmsk r0, 0x1
09: retsp 0x2

\end{lstlisting}
\end{minipage}\hfill
\begin{minipage}{.52\textwidth}
\bgroup
\createlinenumber{7}{7a}
\createlinenumber{8}{7b}
\createlinenumber{14}{14a}
\createlinenumber{15}{14b}

\begin{lstlisting}[language=hcir, frame=tlrb, numbers=left, numberblanklines=false]
fact(R0,R0_3):-	
   entsp(0x2),
   stw(R0,Sp0x1),
   ldw(R1,Sp0x1),
   ldc(R0_1,b0x0),
   lss(R0_2,bR0_1,R1),
   bf(R0_2,0x8),
   fact_aux(R0_2,Sp0x1,R0_3,R1_1).

fact_aux(1,Sp0x1,R0_4,R1):-
   bu(0x0A),
   ldw(R0_1,Sp0x1),
   sub(R0_2,R0_1,0x1),
   bl(fact),
   fact(R0_2,R0_3),
   ldw(R1,Sp0x1),
   mul(R0_4,R1,R0_3),
   retsp(0x2).

fact_aux(0,Sp0x1,R0,R1):-
   mkmsk(R0,0x1),
   retsp(0x2).
\end{lstlisting}
\egroup
\end{minipage}
\vspace{-2.5mm}
\caption{An ISA (factorial) program (left) and its Horn-clause representation (right).}
\label{fig:blocks-transform}
\figend
\end{figure}

Consider the example in Fig.~\ref{fig:blocks-transform} (right).  The
following 
cost equations are set up over the function \emph{fact} that
characterize the energy consumption of the whole function using the
approximation $A$ (e.g., upper/lower)
of each block inferred by the EA, as a function of its input data size
$R0$ (in this case the metric is the integer value of $R0$):
\vspace{-1.5mm}
$$
\begin{array}{rcl}
fact_{e}^A(R0) & = & B1_e^A + fact\_aux_{e}^A(0 \leq R0, R0)  \\
\end{array}
$$
\vspace*{-5mm}
$$
\begin{array}{rcl}
fact\_aux_{e}^A(B, R0)  = 
\left\{
\begin{array}{lr}
B2_e^A+ fact_{e}^A(R0 - 1) 
 \ \  \text{if } B \text{ is {\tt true}}  \\
B3_e^A \quad\quad\quad\quad\quad\quad\quad\ \ \text{if } B \text{ is {\tt false}}
 \
\end{array}
\right.
\end{array}
$$

These inferred recurrence relations/equations are then passed on to a
computer algebra system (e.g., CiaoPP’s internal solver or an external
solver such as Mathematica, both used for the results presented in
this paper) in order to obtain a closed form function for them. If we
assume (for simplicity of exposition) that each basic block has
unitary cost in terms of energy consumption, i.e., $Bi_e = 1$ for all
$i$, we obtain the energy consumed by \texttt{fact} as a function of
its input data size $R0$ as:
$fact_{e}(R0) = R0 + 1$. 

The functions inferred by the static analysis are arithmetic,
(including polynomial, exponential, logarithmic, etc.), and their
arguments (the input data sizes) are natural numbers.
The generic resource analyzer ensures that the inferred bounds are
strict/safe if it is supplied with energy models which provide safe
bounds. 
As mentioned in the introduction,
in~\cite{isa-energy-lopstr13-final-short} we performed a previous
instantiation of such generic analyzer by using the instruction-level
energy model described in~\cite{Kerrison13-short}. However, that model
provides average energy values. As a result, the analysis inferred an
upper-bound energy function for the whole program that was an
approximation of the actual upper bound, that could possibly be below
it.

\secbeg
\section{Experimental Assessment}
\label{sec:experiments}
\secend

In this section we report on an experimental evaluation of our
approach to inferring both upper and lower bounds on the energy
consumed by program executions, given as functions on input data
sizes.

\noindent\textbf{Language and Platform Modeled.}
\label{sec:xmos}
As mentioned before, the experiments have been performed with
XC programs running on the \xmos XS1-L
architecture~\cite{XS1-Architecture}.  Such programs include
typical embedded applications, e.g., signal (audio) processing,
for which the XS1-L architecture was mainly designed.
As also mentioned before, the \xmos XS1 is a cache-less, predictable
architecture by design, 
with a 4-stage pipeline that only permits a single instruction per
thread to be active within the pipeline at the same time,
and thus avoids pipeline hazards.
The particular (development) hardware for which we derive the
branchless-block-level model is a dual-tile board, designed by \xmos,
that contains an XS1-A16-128-FB217~processor.

\noindent\textbf{The Measurement Harness.}
\label{sec:measurement-harness}
In order to take power measurements during execution on real hardware,
record and/or display them in real time, the hardware and software
harness designed by \xmos, as an extension of the \xmos toolchain,
includes:

\begin{itemize}

\item A (hardware) debug adapter (xTAG v3.0) that enables power to be
  measured~\cite{XTAG-3-manual-2015}.
The basic principle consists in placing a small shunt resistor of
$R_{shunt}$ ohm
in series within the supply line.
By measuring
the voltage drop on the shunt $V_{shunt}$, 
the current is calculated as $I_{shunt} = V_{shunt} / R_{shunt}$
(Ohm's law), which is also the current of the power
supply $I_{sup} = I_{shunt}$.
Then the power consumption is estimated as $V_{sup} \times I_{sup}$,
where $V_{sup}$ is the voltage of the power supply.
The xTAG v3.0 adapter has an extra connector that carries the analog
signals required to estimate the power consumption, as explained
above. The measurements regarding these signals are transported to the
host computer over USB using the xSCOPE
interface~\cite{xmos-xscope}. 

\item A (software) tool (\verb+xgdb+, the debugger),
which collects data from the xTAG to be used by the analysis, by
connecting to it over a USB interface (using libusb), and reading both
ordinary xSCOPE traffic and voltage/current measurements.
\end{itemize} 

\subsecbeg
\subsection{Experimental Results and Discussion}
\label{res}
\subsecend

The aim of the experimental evaluation is to perform a first
comparison of the actual upper and lower bounds on energy consumption
measured on the hardware against the respective
bounds obtained by evaluating the functions inferred by our proposed
approach (which depend on input data sizes), for each program
considered and for a range of input data sizes.  For a given input
data size $n$ the actual upper and lower bounds measured on the
hardware where obtained by using data of size $n$ that exhibit the
worst and best cases respectively.

\newcommand{\dbranchno}{n}
\newcommand{\dbranchyes}{y}

The selected benchmarks, which are either iterative or recursive, are
shown in Table~\ref{tab:comparison}. For conciseness, the first column
only shows the names of the programs and the arguments that are
relevant for their energy-bound functions. The \databranching~column
expresses whether a benchmark has data-dependent branching or not
(y/n).  The third
column shows the upper- and
lower-bound energy functions (on input data sizes) inferred by our
approach,
as well as the size metric used. When an input argument (in the first
column) is numeric, its size metric is its actual value (and is
omitted in the third column). Column \blockmodelHW{} shows the
average deviation
of the energy estimations obtained by evaluating such functions,
with respect to the actual bounds measured on the hardware as
explained above. A deviation is positive (resp.\ negative) if the
estimated value is
over (resp.\ under) the actual measurement.

\begin{table}[t]
\centering
\tabbeg
\begin{tabular}{|l|c|p{5.7cm}|p{1.3cm}|}
\hline 
\textbf{Program} & \databranching & {\textbf{Upper/Lower Bounds (nJ)}$\times 10^3$} & \blockmodelHW 
\\ \hline \hline 

$fact(N)$ & 
\multirow{2}{*}{\dbranchno}
  &  $ub=5.1\ N+4.2$      & $+7\%$ \\  
& &$lb=4.1\ N+3.8$& $-11.7\%$  \\ \hline

$fibonacci(N)$ & 
\multirow{2}{*}{\dbranchno}
  &   $ub$\tablefootnote{
The mathematical function $lucas(n)$ satisfies the recurrence relation $lucas(n)=lucas(n-1)+lucas(n-2)$ with $lucas(1)=1$ and $lucas(2)=3$.}$ =5.2\ lucas(N)$
$+6\ fib(N) -6.6$   & $+8.71\%$   \\
& & $lb=4.5\ lucas(N)+5\ fib(N) -4.2$ & $-4.69\%$ \\  \hline

$reverse(A)$ & 
\multirow{2}{*}{\dbranchno} 
 & $ub=3.7\ N+13.3$ (\scriptsize $N =$ length of array $A$) & $+8\%$ \\
 & & $lb=3\ N+12.5$ & $-8.8\%$  \\ \hline

$findMax(A)$  & 
\multirow{2}{*}{\dbranchyes} 
& $ub=5\ N+6.9$ (\scriptsize $N =$ length of array $A$) & $+8.7\%$ \\
& & $lb=3.3\ N+5.6$& $-9.1\%$ \\ \hline

$selectionSort(A)$  & \multirow{2}{*}{\dbranchyes} &
 $ub=30\ N^2 + 41.4\ N+10$ (\scriptsize $N =$ length of array $A$) & $+8.7\%$ \\
& & $lb=16.8\ N^2 + 28.5\ N+8$& $-9.1\%$ \\ \hlinewd{1.1pt} 
 
$fir(N)$ & \multirow{2}{*}{\dbranchyes} & $ub=6\ N+26.4$ & $+8.9\%$   \\
& & $lb=4.8\ N+22.9$ & $-9.7\%$ \\ \hline 

$biquad(N)$ & \multirow{2}{*}{\dbranchyes} & $ub=29.6\ N+10$ & $+9.8\%$  \\
& & $lb=23.5\ N+9$ & $-11.9\%$ \\ \hline 

\end{tabular}
\vspace{1mm}
\caption{Accuracy of upper- and lower-bound estimations.}

\label{tab:comparison}
\tabend
\vspace{-2mm}
\end{table}

The first two benchmarks are small arithmetic programs.  The third
benchmark $reverse(A)$ reverses elements of an input array $A$ of size
$N$.
A sorting algorithm ($selectionsort$) and a simple program for finding
the maximum number in an array ($findMax$) are also included. The
latter, which is also part of the former, is a program where
data-dependent branching can bring significant variations in the
worst- and best-case energy consumption for a given input data size.
We have also studied two audio signal processing benchmarks, 
$biquad$ and $fir$ (Finite Impulse Response), provided by \xmos as
representatives of XS1 application kernels.  Both programs perform
filtering tasks that attenuates or amplifies specific frequency ranges
of a given input signal.

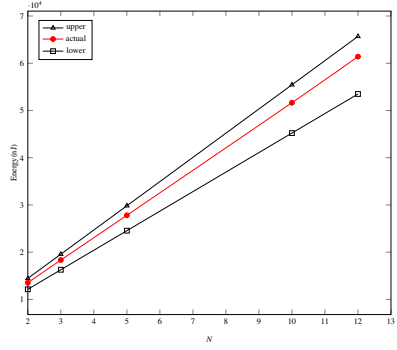
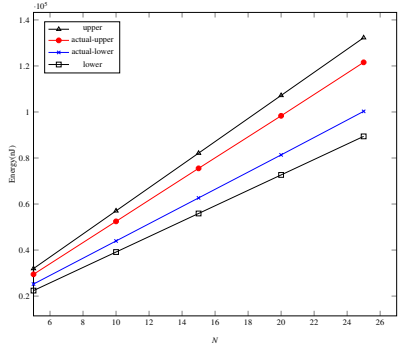
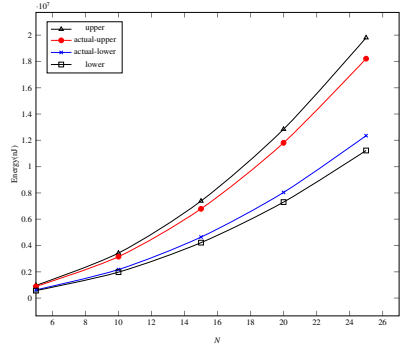
\begin{figure}[ht]
\figbeg\figbeg
 \centering
\subfigure[$fact$.  
]{\label{fig:factGraph}
\begin{tikzpicture}[scale=0.45]
  \begin{axis}[
    legend style={legend pos=north west,font=\scriptsize},
    xlabel=$N$,
    xmin = 2,
    samples at = {2,3,5,10,12},
    ylabel={Energy(nJ)},
    tick label style={font=\scriptsize},
    label style={font=\scriptsize},
    smooth,    
    width=\linewidth,
	y label style={at={(axis description  cs:+.1, .5)},  anchor=south}
  ] 
    \addplot[smooth,mark=triangle,black] {(5122628*x+4210630)/1000}; 
    \addlegendentry{upper}

        \addplot[smooth,mark=*,red] coordinates {
	  (2,   13534.10551)
 	  (3,   18341.04856)
	  (5,   27808.21706)
	  (10,   51636.375)
	  (12,   61410.44654)
	};
	
    \addlegendentry{actual}

    \addplot[smooth,mark=square,black] {(4134150*x+3879193)/1000};
    \addlegendentry{lower}

  \end{axis}
\end{tikzpicture}}

\subfigure[$findMax$. 
]{\label{fig:findMax}
\begin{tikzpicture}[scale=0.45]
  \begin{axis}[
    legend style={legend pos=north west,font=\scriptsize},
    xlabel=$N$,
    xmin = 5,
    samples at = {5,10,15,20,25},
    ylabel={Energy(nJ)},
    tick label style={font=\scriptsize},
    label style={font=\scriptsize},
    smooth,    
    width=\linewidth,
    y label style={at={(axis description  cs:+.08, .5)},  anchor=south}
  ] 
    \addplot[smooth,mark=triangle,black] {(5015535*x+6882011)/1000}; 
    \addlegendentry{upper}

        \addplot[smooth,mark=*,red] coordinates {
	  (5,   29461.71413)
 	  (10,   52466.924)
	  (15,   75505.4205)
	  (20,   98332.7395)
	  (25,   121567.227)
	};
	
    \addlegendentry{actual-upper}
    
        \addplot[smooth,mark=x,blue] coordinates {
	  (5,   25230.56763)
 	  (10,   43949.277)
	  (15,   62658.0315)
	  (20,   81354.0915)
	  (25,   100269.0045)
	};
	
    \addlegendentry{actual-lower}

    \addplot[smooth,mark=square,black] {(3350309*x+5640360)/1000};
    \addlegendentry{lower}

  \end{axis}
\end{tikzpicture} 
}
\subfigure[$selectionsort$. 
]{\label{fig:selsort}
\begin{tikzpicture}[scale=0.45]
  \begin{axis}[
    legend style={legend pos=north west,font=\scriptsize},
    xlabel=$N$,
    xmin = 5,
    samples at = {5,10,15,20,25},
    ylabel={Energy(nJ)},
    tick label style={font=\scriptsize},
    label style={font=\scriptsize},
    smooth,    
    width=\linewidth,
    y label style={at={(axis description  cs:+.08, .5)},  anchor=south}
  ] 
    \addplot[smooth,mark=triangle,black] {(30015730*x*x+41425609*x+10053218)/1000}; 
    \addlegendentry{upper}

        \addplot[smooth,mark=*,red] coordinates {
	  (5,   889700)
 	  (10,   3150200)
	  (15,   6790700)
	  (20,   11811200)
	  (25,   18211700)
	};
	
    \addlegendentry{actual-upper}
    
        \addplot[smooth,mark=x,blue] coordinates {
	  (5,   628300)
 	  (10,   2172800)
	  (15,  4642300)
	  (20,   8036800)
	  (25,   12356300)
	};
	
    \addlegendentry{actual-lower}

    \addplot[smooth,mark=square,black] {(16810328*x*x+28523082*x+8142371)/1000};
    \addlegendentry{lower}
    
  \end{axis}
\end{tikzpicture} 
}
\figend
\caption{Estimated energy upper/lower bounds vs.\ actual measurements.}
\figend\figend
\end{figure}

Figure~\ref{fig:factGraph} depicts the upper- and lower-bound energy
functions inferred by the analysis, as well as the actual bounds
measured on the hardware for the $fact(N)$
program (taking different values of $N$).
In this case, both the actual upper- and lower-bounds coincide, as
shown by the middle curve (in red), which plots the actual
measurements on the hardware.
It can be observed that the values of the upper-bound function
inferred by the static analysis supplied with the model obtained by
the EA always over-approximate the actual hardware measurements (by
7\%, as given by Table~\ref{tab:comparison}), whereas the lower-bound
values under-approximate the actual measurements (by 11.7\%).

Similarly, the $findMax$ benchmark
is shown in Figure~\ref{fig:findMax}. Unlike $fact$, the actual upper-
and lower-bound functions of $findMax$, depending on input arrays of
length $N$, do not coincide,
due to the data-dependent
branching. The actual energy consumption of $findMax$ not only depends
on the length of the input array, but also on its contents, and thus
cannot be captured exactly by a function that depends on data sizes
only (i.e., by abstracting the data by their sizes). A call to
$findMax$ with a sorted array in ascending order (of a given length
$N$) will discover a new \emph{max} element in each iteration, and
hence update the current \emph{max} variable, resulting in the actual
upper-bound (i.e., worst case of the algorithm). In contrast, if the
array is sorted in descending order, the algorithm will find the
\emph{max} element in the first iteration, and the rest of the
iterations will never update the current \emph{max} variable,
resulting in the actual lower-bound (i.e., best case).
Thus, Figure~\ref{fig:findMax} depicts four curves: the upper- and
lower-bound energy functions inferred by our approach for $findMax$,
as well as the two actual energy bound curves
measured on the hardware. The former are obtained by evaluating the
energy functions in Table~\ref{tab:inaccuracy_src}, for different
array-lengths $N$, as before. The latter are obtained with actual
arrays of length $N$ that give the worst and best cases, as explained
above.  Note that it is not always trivial to find data that exhibit
program worst and best case behaviors.
Table~\ref{tab:comparison} shows that the inferred upper-
(resp. lower-) bounds over- (resp. under-) approximate the actual
upper- (resp. lower-) bounds measured on the hardware by 8.7\%
(resp. 9.1\%).  Figure~\ref{fig:selsort} for $selectionsort$ shows 
a similar behavior but with quadratic bounds.

The inaccuracies in the energy estimations of our technique come
mainly from two sources: the modeling, which assigns an energy value
to each basic block as described in Section~\ref{sec:energymodel}, and
the static analysis, described in Section~\ref{sec:static}, which
estimates the number of times that the basic blocks are executed
depending on the input data sizes, and hence, the energy consumption
of the whole program.  Table~\ref{tab:inaccuracy_src} shows part of
the results of our study in order to quantify the inaccuracy
originating from those sources.  Different executions of the $findMax$
benchmark are shown for different input arrays of length \textbf{N}
(Column \textbf{N}). The table is divided into two parts. The first part
uses randomly generated input arrays of length \textbf{N}, while the
second part (three lower rows) uses input arrays that cause the worst-
and best-case energy consumption. Column \textbf{Cost App} indicates
the type of approximation of the automatically inferred energy
functions:
upper bound (\textbf{U}) and lower bound (\textbf{L}). Such energy
functions 
are shown in Table~\ref{tab:comparison}.
We have then compared the energy consumption estimations obtained by
evaluating the energy function (Column \textbf{Est}) with the observed
energy consumption of the hardware measurements (Column \textbf{Obs}).
Column \textbf{D} shows the relative harmonic difference between the
estimated and the observed energy consumption, given by the formula:

\vspace{-3mm}
$$rel\_harmonic\_diff(Est, Obs) = \frac{(Est - Obs) \times (\frac{1}{Est} + \frac{1}{Obs})}{2}$$
\vspace{-5mm}

\begin{table}[t]
\centering
\tabbeg
\setlength{\tabcolsep}{1em}
\begin{tabular}{|c|c|r|r|r|r|r|}

\hline 
\multirow{2}{*}{\textbf{N}} &\textbf{Cost}  &\multicolumn{3}{|c|}{\textbf{Energy(nJ)}$\times 10^3$} & \multirow{2}{*}{\textbf{D \%}} & \multirow{2}{*}{\textbf{PrD \%}}
\\ \cline{3-5}
&\textbf{App}&{\textbf{Est}} & \textbf{Prof} & \textbf{Obs}&& \\
\hline \hline 

\multicolumn{7}{|c|}{Random array data} \\ \hline

\multirow{2}{*}{\texttt{5}}  
  &  L & 22.3 & 24.9 & \multirow{2}{*}{27.3} & -20.1 & -9.2 \\
  &  U & 31.9 & 30.2 &  & 15.6 & 10
\\ \hline
\multirow{2}{*}{\texttt{15}}  
  &  L & 55.9 & 61.8 & \multirow{2}{*}{69.1} & -17 & -11 \\
  &  U & 82.1 & 75.1 &  & 21 & 8.3
\\  \hline
\multirow{2}{*}{\texttt{25}}  
  &  L & 89.4  & 99.6  & \multirow{2}{*}{110.9} & -17.6 & -10.7 \\
  &  U & 132.2 & 120.8 &  & 21.7 & 8.5 \\ \hline
  
\multicolumn{7}{|c|}{Actual worst- and best-case array data} \\ \hline

\multirow{2}{*}{\texttt{5}}  
  &  L & 22.3 & 22.3 & 25.2 & -12.2 & -12.2 \\
  &  U & 31.9 & 31.9 & 29.4 & 8.1 & 8.1
\\ \hline
\multirow{2}{*}{\texttt{15}}  
  &  L & 55.9 & 55.9 & 62.6 & -11.3 & -11.3 \\
  &  U & 82.1 & 82.1 & 75.5 & 8.3 & 8.3
\\  \hline
\multirow{2}{*}{\texttt{25}}  
  &  L & 89.4  & 89.4  & 100.2 & -11.4 & -11.4 \\
  &  U & 132.2 & 132.2 & 121.5 &  8.4 & 8.4

\\ \hline
\end{tabular}
\vspace{1mm}
\caption{Source of inaccuracies in $findMax$ prediction: analysis vs.\ modeling.}
\label{tab:inaccuracy_src}
\tabend 
\end{table}

Column \textbf{Prof} shows the result of estimating the energy
consumption using the energy model and assuming that the static
analysis was perfect and estimated
the exact number of times that the basic blocks were executed. This
obviously represents the case in which all loss of accuracy must be
attributed
to the energy model.  
The values in Column \textbf{Prof} have been obtained  
by profiling actual executions of the program with the concrete input
arrays, where the profiler has been instrumented to record the number
of times each basic block is executed. The energy consumption of the
program is then obtained by multiplying such numbers by the values
provided by the energy model for each basic block, and adding all of
them.  Column \textbf{PrD} represents the inaccuracy due to the energy
modeling of basic blocks using the EA, which has been quantified as
the relative harmonic difference between \textbf{Prof} and the
observed energy consumption \textbf{Obs}. The difference between
\textbf{D} and \textbf{PrD} represents the inaccuracy due to
the static analysis.

Although the first part of the table, using random data, may give the
impression that both the static analysis and the energy modeling
contribute to the inaccuracy of the energy estimation of the whole
program, the second (lower) part of the table indicates that the
inaccuracy only comes from the energy modeling.  This is because in
the lower part the comparison was performed with input arrays that
make $findMax$ exhibit its actual upper- and lower-bounds (depending
on the length of the array).
In this case, Columns \textbf{Est} and \textbf{Prof} show the same
values, which means
that there was no inaccuracy due to the static analysis (regarding the
inference of the \emph{actual upper- and lower-bound functions}), and
that the overall inaccuracy is due to the over- and
under-approximation in the EA to model energy consumption of each
basic block.

Table~\ref{tab:inaccuracy_src_reverse} shows a similar experiment for
the $reverse$ program, which has no data-dependent branching.  Since
the number of operations performed by
$reverse$ is actually a function of the length of its input array (not
of its contents), Columns \textbf{Est} and \textbf{Prof} show the same
values for random data (unlike for $findMax$), which means that no
inaccuracy comes from the static analysis part.

\begin{table}[t]
\centering
\tabbeg
\setlength{\tabcolsep}{1em}
\begin{tabular}{|c|c|r|r|r|r|r|}
\hline 
\multirow{2}{*}{\textbf{N}} &\textbf{Cost}  &\multicolumn{3}{|c|}{\textbf{Energy(nJ)}$\times 10^3$} & \multirow{2}{*}{\textbf{D \%}} & \multirow{2}{*}{\textbf{PrD \%}}
\\ \cline{3-5}
&\textbf{App}&{\textbf{Est}} & \textbf{Prof} & \textbf{Obs}&& \\
\hline \hline 
\multicolumn{7}{|c|}{Random array data} \\ \hline
\multirow{2}{*}{\texttt{5}}  
  &  L & 28 & 28 & \multirow{2}{*}{29} & -3.5 & -3.5 \\
  &  U & 31.8 & 31.8 &  & 9.2 & 9.2
\\ \hline
\multirow{2}{*}{\texttt{15}}  
  &  L & 59 & 59 & \multirow{2}{*}{64} & -8.1 & -8.1 \\
  &  U & 68.8 & 68.8 &  & 7.2 & 7.2
\\  \hline
\multirow{2}{*}{\texttt{25}}  
  &  L & 90  & 90  & \multirow{2}{*}{98} & -8.5 & -8.5 \\
  &  U & 105.8 & 105.8 &  & 7.7 & 7.7 \\ \hline
\end{tabular}
\vspace{1mm}
\caption{Source of inaccuracies in $reverse$ prediction: analysis vs.\ modeling.}
\label{tab:inaccuracy_src_reverse}
\tabend 
\end{table}

Regarding the time taken by the EA, it can vary 
depending on the parameters it
is initialized with, as well as the initial population. This 
population is different every time the EA is initiated, except for a
fixed number of individuals that represent corner cases.  In the
experiments, the EA 
is run for up to a maximum of 20
generations, and is stopped
when the fitness value does not improve for four consecutive
generations. In all the experiments the $biquad$ benchmark took
the most time (a maximum time of 230 minutes) for maximizing the
energy consumption. In contrast, the \emph{fact} benchmark took the
least time (a maximum time of 121 minutes). The times remained within
the 150-200 minutes range on average. Time speed-ups were also
achieved by reusing the EA results for sequences of instructions that
were already processed in a previous benchmark (e.g., return blocks,
loop header blocks, etc.). 
This makes us believe that our approach could be used in practice in
an iterative development process, where the developer gets feedback
from our tool and modifies the program in order to reduce its energy
consumption. The first time the EA is run would take the highest 
time, since it would have to determine the energy consumption of all
the program blocks. After a focused modification of the program that
only affects a 
small number of blocks, most of the results from the previous run
could be reused, so that the EA would run much faster during this
development process. In other words, the EA processing can easily be
made incremental. 

The static analysis, on the other hand, is quite efficient, with
analysis times of about 4 to 5 seconds on average, despite the naive
implementation of the interface with external recurrence equation
solvers, which can be improved significantly. 

\secbeg
\section{Related Work}
\label{sec:related-work}
\secend

Static
analysis of the energy consumed by program executions has received
relatively little attention until recently.  An analysis of Java
bytecode programs that inferred upper-bounds on energy consumption as
functions on input data sizes was proposed
in~\cite{NMHLFM08-tooshort}, where the Jimple (a typed three-address
code) representation of Java bytecode was transformed into Horn
Clauses, and a simple energy model at the Java bytecode
\level~\cite{LL07} was used.  However the energy model used average
estimations of the Java opcodes, which are not suitable for
verification applications.
Furthermore, this work did not compare the results with actual,
measured energy consumption.  As already mentioned, a similar approach
was proposed in~\cite{isa-energy-lopstr13-final-short} for the
analysis of XC programs. However, it used an ISA-level model that also
provided average energy values, which implied the same problem for
verification.
Other approaches to static analysis based on the transformation of the
analyzed code into another (intermediate) representation have been
proposed for analyzing low-level languages~\cite{HGScam06-short} and
Java (by means of a transformation into Java
bytecode)~\cite{jvm-cost-esop-short}. In~\cite{jvm-cost-esop-short},
cost relations are inferred directly for these bytecode programs,
whereas in~\cite{NMHLFM08-tooshort} the bytecode is first transformed
into Horn Clauses~\cite{decomp-oo-prolog-lopstr07-shortest}.

Other work has taken as its starting point techniques referred to
generally
as \emph{WCET} (Worst Case Execution Time Analyses), which have been
applied, usually for imperative languages, in different application
domains (see e.g.,~\cite{DBLP:journals/tecs/WilhelmEEHTWBFHMMPPSS08}
and its references).  These techniques generally require the
programmer to bound the number of iterations of loops,
and then apply an Implicit Path Enumeration technique to identify the
path of maximal consumption in the control flow graph of the resulting
loop-less program. This approach has inspired some worst case energy
analyses, such 
as~\cite{Jayaseelan_2006_EWE-short}.
It distinguishes instruction-specific (not proportional to time, but
to data) from pipeline-specific (roughly proportional to time) energy
consumption. The approach also takes into account complex issues such
as branch prediction and cache misses. However, they rely on the user
to identify the input which will trigger the maximal energy
consumption.
In~\cite{Wagemann-2015-WCEC} the same approach is
further
refined for estimating \emph{hard} (i.e., over-approximated) energy
bounds. The main novelty of this work consists in introducing relative
energy models (implemented at the LLVM level in this case), where the
energy of instructions is given \emph{in relation to each other}
(e.g., if we assume that all the instructions have relative energy 1,
this means that they all have the same absolute energy), which does
not depend on the specific hardware, but can be applied for all the
platforms where a mapping between LLVM and low-level 
ISA 
instructions exists. On the other hand, in
situations when the
energy bounds are not \emph{hard} (i.e., the application allows their
violation) they use a genetic algorithm to obtain an
under-approximation of the energy bounds. However, this approach loses
accuracy
when there are data-dependent branches present in the program, since
different inputs can lead to the execution of different sets of
instructions.
A similar approach is used in~\cite{pallister2015data} to find the
worst-case energy consumption of two benchmarks using a genetic
algorithm.
In contrast to our approach, the evolutionary algorithm is applied to
whole programs, which are required to not have any data-dependent branching.
The authors further introduce probability distributions for the
transition costs among pairs of independent instructions, which can 
then be convolved to give a probability distribution of the energy for a
sequence of instructions.

In contrast to the work presented here and
in~\cite{estim-exec-time-ppdp08-short}, all these WCET-style
methods (either for execution time or energy) do not infer cost
functions on input data sizes but rather absolute maximum values, and,
as mentioned before, they generally require the manual annotation of
all loops to express an upper bound on the number of iterations, which
can be tedious (or impossible). Loop bound inference techniques can
also be applied but require that all loop counts can be resolved.
All of this essentially reduces the case to that of programs with no
loops.
Another alternative approach to WCET-style methods was presented
in~\cite{Herrmann-WCET-2007-short}.  It is based on the idea of
amortization, which allows inferring more accurate yet safe upper
bounds by averaging the worst execution time of operations over time.
It was applied to a functional language, but the approach is in
theory generally applicable and could in principle be
adapted to inferring energy usage.

\secbeg
\section{Conclusions}
\label{sec:conclusion}
\secend

We have proposed an approach for inferring parametric upper and lower
bounds on the energy consumption of a program using a combination of
static and dynamic techniques.  The dynamic technique, based on an
evolutionary algorithm, is used to determine the maximum/minimum
energy consumption of the basic blocks in the program.  Such blocks
contain multiple instructions, which allows this phase to capture
inter-instruction dependencies. Moreover, the basic blocks are
branchless, which makes the evolutionary algorithm approach quite
practical and efficient, and the energy values inferred by it are
accurate, since no control flow-related variations occur.
A static analysis is then used to combine the energy values obtained
for the blocks according to the program control flow, and produce
parametric energy consumption bounds of the whole program that depend
on input data sizes.  We also carried out an experimental study to
validate the upper and lower bounds on a set of benchmarks. The
results support our hypothesis that the bounds inferred by our
approach are indeed safe and quite accurate, and the technique
practical for its application to energy verification and
optimization.

\vspace{2mm}\noindent\textbf{Acknowledgments.} This research has
received funding from the European Union 7th Framework Program
agreement no 318337, ENTRA, Spanish MINECO TIN2012-39391
\emph{StrongSoft} and TIN2015-67522-C3-1-R \emph{TRACES} projects, and
the Madrid M141047003 \emph{N-GREENS} program.  We also thank Henk
Muller, Principal Technologist, \xmos, for his help with the
measurement boards, evaluation platform, benchmarks, and overall
support.

\secbeg
\bibliographystyle{abbrv}

\vfill
\end{document}